
\magnification=\magstep1          \def\sp{\,\,\,} \overfullrule=0pt
 \def\h{h^\vee} \def\c{\chi} \def\C{\chi^*}  \def\v{\vartheta}
\def\T{{\cal T}}   
\def\l{\Lambda}   \def\la{\lambda}    \def\Z{{\bf Z}}
\def\u{\tau}       \def\equi{\,{\buildrel \rm def \over =}\,}
\def\eg{{\it e.g.}$\sp$} \def\ie{{\it i.e.}$\sp$}
 \def\g{{\hat g}}

{\nopagenumbers
\rightline{August, 1992}
\bigskip \bigskip
\centerline{{\bf WZW Commutants, Lattices, and}}\bigskip
\centerline{{\bf Level 1 Partition Functions}}
\bigskip \bigskip
\centerline{Terry Gannon}
\centerline{{\it Mathematics Department, Carleton University}}
\centerline{{\it Ottawa, Ont., Canada K1S 5B6}}\bigskip \bigskip \bigskip
A natural first step in the classification of all `physical' modular
invariant partition functions $\sum N_{LR}\,\c_L\,\C_R$ lies in understanding
the commutant of the modular matrices $S$ and $T$. We begin this paper
extending the work of Bauer and Itzykson on the commutant from the $SU(N)$
 case they
consider to the case where the underlying algebra is any semi-simple
Lie algebra (and the levels are arbitrary). We then use this
 analysis to show that the partition functions
associated with even self-dual lattices span the commutant. This proves that
the lattice method due to Roberts and Terao, and Warner, will succeed in
generating all partition functions. We then make some general remarks
concerning
certain properties of the coefficient matrices $N_{LR}$, and use those
to explicitly find
all level 1 partition functions corresponding to the algebras $B_n$, $C_n$,
$D_n$, and the 5 exceptionals. Previously,
only those associated to $A_n$ seemed to be generally known.
\vfill \eject} \pageno=1

\centerline{{\bf 1. Introduction}} \bigskip

The relevance of two-dimensional conformal field theories [1] to the study of
string theories and second order phase transistions in statistical systems
is well-known, hence the considerable attention devoted in the literature
to their classification.  An important subclass of these are the
Wess-Zumino-Witten models [2].
The partition function of one
associated with (untwisted affine) Kac-Moody algebra $\g=g^{(1)}$ and level
$k$ [3,4] can be written in the following way:
$$Z(z_Lz_R|\u)=\sum N_{\lambda_L \lambda_R} \,\c_{ \lambda_L}
(z_L,\u)\,\c_{\lambda_R}(z_R,\u)^*. \eqno(1.1)$$
$\c_{\lambda}$
 is the {\it normalized character} (see [3]) of the representation of
 $\g$ with (horizontal) highest weight $\lambda$.
The (finite) sum in eq.(1.1) is over all level $k$ highest weights $\la_L,$
$\la_R$. The coefficients $N_{\la_L\la_R}$ are numbers (multiplicities).

Most writers consider the {\it restricted} partition functions $Z(\u)\equi
Z(0,0|\u)$ (however in [5] it is argued that for $c\ge 1$ the complex vectors
$z$ should be
retained because the restricted partition functions cannot carry enough
information to specify the theory). In this paper we will retain the vectors
$z$. Of course the restricted partition functions can be recovered at the
end by substituting $z=0$.

We will restrict our attention
to the case where the right- and left-moving sectors correspond
to the same algebra $\g$ and level $k$.
For a discussion of the {\it heterotic} case, where $\g_L,k_L\ne \g_R,k_R$,
see [6].

There are three properties the sum in eq.(1.1) must satisfy in order
to be the partition function of a sensible conformal field theory:
\item{(P1)} it must be {\it modular invariant}. This is equivalent to
the two conditions:
$$\eqalignno{&Z(z_Lz_R|\u+1)=Z(z_Lz_R|\u), &(1.2a)\cr
\exp[-k\pi i(z_L^2/\u-z_R^{*2}/\u^*)]\,
&Z(z_L/\u, z_R/\u|-1/\u)=Z(z_L z_R|\u); &(1.2b)\cr}$$

\item{(P2)} the coefficients $N_{ \lambda_L \lambda_R}$
in (1.1) must be {\it non-negative integers}; and

\item{(P3)} $N_{00}=1$ (the zero vector here is the highest weight of
the singlet representation of level $k$).

If a function $Z$ in (1.1) satisfies (P1), we will call it an {\it invariant};
if in addition each $N_{\la_L,\la_R}\ge 0$, we shall call it a {\it positive
invariant};
 and if it satisfies (P1), (P2), and (P3) we will call it a {\it physical
invariant}. This paper is concerned with the problem of finding all
physical invariants corresponding to a given choice of algebra and level.
It will not address the question of which of these physical invariants are
actually realized by a well-defined theory.

The problem of classifying all physical invariants is a difficult one.
Several examples of these invariants are known. The usual techniques for
finding these include conformal embeddings [7], simple currents [8] and
outer automorphisms of the Kac-Moody algebra [9]. The pattern seems
to be that for a fixed algebra $\g$, there are a finite number of infinite
 series ${\cal A}_k$,
${\cal D}_k$, etc. of physical invariants, each defined for all levels $k$
lying on certain arithmetic sequences,
together with some
{\it exceptional} invariants ${\cal E}_k$, defined only for finitely
many $k$. The most famous example is the A-D-E classification for
$\g=A_1^{(1)}$ [10]. (In Sec.5 we make a small step towards establishing
this general pattern, by proving that for each choice of algebras and levels,
there are only finitely many physical invariants.)

In spite of the large numbers of known physical invariants, there are very
few {\it completeness proofs} which determine {\it all} physical invariants
belonging to a certain class.
The most significant example is the $A_1^{(1)}$ completeness proof [10].
Another one concerns the level 1 $A_n^{(1)}$ physical
invariants [11].
In ref.[8] it was remarked that for level 1 simply-laced $\g$, the method
of simple currents reduces to the bosonic lattice construction, and for
this reason conjectured their list of physical invariants for $k=1$, $\g=
D_n^{(1)}$, $E_6^{(1)}$, $E_7^{(1)}$ and $E_8^{(1)}$ was complete.
We will prove this conjecture in Sec.5, and complete the
search for all level 1 physical invariants by finding all corresponding
to the non-simply-laced $\g$ as well. These $k=1$ physical invariants are
explicitly listed in Thm.5.

A frustrating feature of these completeness proofs is that they tend to be
primarily
number-theoretic, unlike the more algebraic techniques for finding these
invariants. Related to this is that, whereas these algebraic techniques
provide a very elegant derivation for several of these invariants,
some of the exceptional invariants
are difficult to derive in these ways.
For these reasons, a method recently given by Warner [12] and, independently,
by Roberts and Terao [13], holds much promise. They propose to generate these
invariants by using the Weyl-Kac formula and theta functions associated with
even self-dual
lattices. Their method will be briefly described below. This method
is more number-theoretic than algebraic. Moreover, all invariants, exceptional
as well as those lying in infinite series, are treated on an equal footing.
At least for small levels and algebras, it is quite practical. We will
prove that this lattice method succeeds in generating all invariants --- in
fact, a small subclass ${\cal L}_*$ of the even self-dual lattices suffice to
span all invariants (see Cor.2).

In Sec.2 we will introduce the notation and terminology used in the later
sections. In Sec.3 we generalize the arguments of [14] and find a basis of
the commutant corresponding to any (semi-)simple algebra and levels.
In Sec.4 we will briefly describe the
Roberts-Terao-Warner lattice method, and then prove its completeness. At the
beginning of Sec.5
we include a few general comments and useful tools concerning physical
invariants and their ``shape''. We apply these later in the section to
find a complete list of all level 1 physical invariants
corresponding to $\g=B_n^{(1)}$, $C_n^{(1)}$, $D_n^{(1)}$, $E_6^{(1)}$,
 $E_7^{(1)},$ $E_8^{(1)},$ $F_4^{(1)}$ and $G_2^{(1)}$.

The essence of the proof in Sec.4 that lattice partition functions span
the commutant lies in
the observation that the integral basis found in [14] for $A_n^{(1)}$
consists essentially of
lattice partition functions. All that remains in Secs.3 and 4 then is
the reasonably straightforward generalization of this argument to any
semi-simple algebra. Sec.5 however is independent of this analysis.
The commutant is not
only a vector space (this is what [14] is exclusively concerned with),
but it also
has a much richer algebraic structure. It is difficult to imagine that this
additional structure will not also be very valuable to exploit. Sec.5 is a
preliminary attempt in that direction.

\bigskip \bigskip \centerline{{\bf 2. Notation and terminology}}
\bigskip

Before we begin the main body of this paper, it is necessary to establish
some notation and terminology, some of which is non-standard.
For a much more complete description of Kac-Moody algebras,
see \eg [3,4].

Let $g$ be any simple finite-dimensional Lie algebra. Let $M=M_g$ be its
{\it coroot lattice}. Then the dual lattice $M^*$ will be spanned by the
fundamental weights $\beta_1,\ldots,\beta_n$, where $n$ is the rank of
$g$ and the dimension of $M$ and $M^*$.
Let $\rho=\sum \beta_i$.

Let $\g=g^{(1)}$ denote the untwisted affine Kac-Moody algebra corresponding
 to $g$.
An integrable irreducible  representation is given by a positive integer
$k$ (called the {\it level}) and a {\it highest weight vector} $\la\in M^*$.
The set of all possible highest weights corresponding to  level $k$
representations will be called $P_{+}(g,k)$, and is defined by
$$P_{+}(g,k)=\bigl\{\sum_{i=1}^n \ell_i \beta_i \,|\, 0\leq \ell_i,\, \ell_i
\in\Z,\, \sum \ell_i a_i^\vee \leq k\bigr\},\eqno(2.1)$$
where the numbers $a_i^\vee$ are called the {\it colabels} of $g$. The
number $1+\sum a_i^\vee$ is denoted $\h$ and is called the
{\it dual Coxeter number}.

The relevant facts about lattices that we require can be found \eg in [15].
By $M^{(\ell)}$ we mean the scaled lattice $\sqrt{\ell}M$. The {\it theta
series} of a translate $v+\l$ of any Euclidean lattice $\l$ is defined to
be
$$\Theta\bigl(v+\l\bigr)(z|\u)\equi \sum_{x\in v+\l}\exp[\pi i \u x^2
+2\pi iz\cdot x],\eqno(2.2)$$
where $\u\in {\bf C}$ satisfies Im $\u>0$, and where the complex vector $z$
lies in the {\it complexification} ${\bf C}\otimes \l\equi \{\sum_i
c_i\cdot x_i\,|\,c_i\in {\bf C},\sp x_i\in \l\}$.

The {\it Weyl-Kac character formula} gives us a convenient expression for
the (normalized) character $\c^{g,k}_\la$ of the representation of $\g$
with level $k$ and highest weight $\la\in P_{+}(g,k)$:
$$\eqalignno{\c_\lambda^{g,k}(z,\u)=&{\sum_{w\in W(g)} \epsilon(w) \,
\Theta\bigl({\lambda+\rho\over \sqrt{k+\h}}+M^{(k+\h)}\bigr)(\sqrt{k+\h}
w(z)|\u)
\over D_g(z|\u)},&(2.3a)\cr D_g(z|\u)\equi &\sum_{w\in W(g)}\epsilon(w)\,
\Theta\bigl( {\rho\over\sqrt{\h}}+M^{(\h)}\bigr)(\sqrt{\h}w(z)|\u).
&(2.3b)\cr}$$
Here, $W(g)$ is the (finite) Weyl group of $g$ and $\epsilon(w)=$det $w\in
\{\pm 1\}$, and $z\in {\bf C}\otimes M$.

By the {\it Weyl-folded commutant} $\Omega_W(g,k)$ we mean the (complex)
space of all functions
$$Z(z_Lz_R|\u)=\sum_{\la,\la'\in P_{+}(g,k)} N_{\la\la'} \,\c_\la^{g,k}
(z_L,\u)\,\c_{\la'}^{g,k}(z_R,\u)^*\eqno(2.4a)$$
invariant under the modular group, \ie those $Z$ in (2.4$a$) satisfying
eqs.(1.2). It is not hard to show that two functions $Z$ and $Z'$ are
equal iff their coefficient matrices $N$ and $N'$ are equal; we will use
the invariant $Z$ interchangeably with its matrix $N$. No confusion should
result.

Our task in the next two sections of this paper is to understand that
commutant. A convenient way to get at it is through a closely related
space, which we will call the {\it theta-commutant} $\Omega_{th}(g,k)$.
It consists of all modular invariant functions
$$\eqalignno{Z(z_Lz_R|\u)=&\sum_{\la,\la'\in \l^*/\l} N_{\la\la'} \,t_\la^{g,k}
(z_L,\u)\,t_{\la'}^{g,k}(z_R,\u)^*,&(2.4b)\cr
t_\la^{g,k}(z,\u)=&{\Theta\bigl(\la+\l\bigr)(\sqrt{k+\h}z|\u)\over D_g(z|\u)},
 &(2.4c)\cr}$$
where $\l=M^{(\h+k)}$. Here also we have $Z=Z'$ iff $N=N'$. The Weyl-Kac
formula (2.3) tells us that by expanding its numerator, any
$Z\in\Omega_W(g,k)$ can also be thought of as
lying in $\Omega_{th}(g,k)$. Moreover, we can use (2.3$a$) to define
$\c_\la^{g,k}$ for {\it all} $\la\in M^*$. Then we learn from [3] that
either
$$\c_\la^{g,k}(z,\u)=0 \eqno(2.5a)$$
for all $z,\u$, or there exists a unique $\epsilon\in \{\pm 1\}$ and
$\mu\in P_{+}(g,k)$ such that
$$\c_\la^{g,k}(z,\u)=\epsilon\,\c_\mu^{g,k}(z,\u)\eqno(2.5b)$$
for all $z,\u$. Thus (2.3) also
defines the process (which we shall call {\it Weyl-folding}) by which
an element of $\Omega_{th}$ can be associated with an element of $\Omega_W$
(see \eg (4.1$d$)).
Note that the theta-commutant $\Omega_{th}$ is isomorphic to
the Hilbert space $E$ considered in [14].

The functions $\c_\la$ and $t_\la$ behave quite nicely under the modular
transformations $\u\rightarrow \u+1$ and $\u\rightarrow -1/\u$:
$$\eqalignno{\c_\la^{g,k}(z,\u+1)=&\sum_{\la'\in P_{+}(g,k)}\big(T^W(g,k)
\big)_{\la\la'}\,\c_{\la'}^{g,k}(z,\u) \sp{\rm where}&(2.6a)\cr
\big(T^W(g,k)\big)_{\la\la'}=&\exp[\pi i {(\la+\rho)^2\over \h+k}-\pi i
{\rho^2\over \h}]\, \delta_{\la\la'}; &(2.6b)\cr
\exp[-k\pi iz^2/\u]\,\,
\c_\la^{g,k}(z/\u,-1/\u)=&\sum_{\la'\in P_{++}(g,k)}\big(S^W(g,k)
\big)_{\la\la'}\,\c_{\la'}^{g,k}(z,\u) \sp{\rm where}&(2.6c)\cr
\big(S^W(g,k)\big)_{\la\la'}={i^{\|\Delta_+\|}\over (\h+k)^{n/2}\sqrt{|M|}}&
\sum_{w\in W(g)}\epsilon(w)\exp[-2\pi i {w(\la'+\rho)\cdot(\la+\rho)\over
\h+k}]
; &(2.6d)\cr
t_\la^{g,k}(z,\u+1)=&\sum_{\la'\in \l^*/\l}\big(T^{th}(g,k)
\big)_{\la\la'}\,t_{\la'}^{g,k}(z,\u),\sp \sp{\rm where}&(2.6e)\cr
\big(T^{th}(g,k)\big)_{\la\la'}=&\exp[\pi i \la^2-\pi i
{\rho^2\over \h}]\, \delta_{\la\la'}; &(2.6f)\cr
\exp[-k\pi iz^2/\u]\,\,
t_\la^{g,k}(z/\u,-1/\u)=&\sum_{\la'\in \l^*/\l}\big(S^{th}(g,k)
\big)_{\la\la'}\,t_{\la'}^{g,k}(z,\u),\sp \sp{\rm where}&(2.6g)\cr
\big(S^{th}(g,k)\big)_{\la\la'}=&{i^{\|\Delta_+\|}\over (\h+k)^{n/2}\sqrt{|M|}}
\exp[-2\pi i \la'\cdot \la].&(2.6h)\cr}$$
In these equations, $\|\Delta_+\|$ denotes the number of positive roots
of $g$, and $|M|$ denotes the {\it determinant} of the lattice $M$. As
before, $\l=M^{(\h+k)}$ and $n$ is the rank of $g$.

The matrices $T^W(g,k)$, $S^W(g,k)$, $T^{th}(g,k)$ and $S^{th}(g,k)$ are
unitary and symmetric. One of the main reasons $\Omega_W$ will be studied
indirectly through $\Omega_{th}$ is that the matrix $S^{th}$ is simpler
than $S^W$.

Note that $Z^W=\sum N^W_{\la\la'}\,\c_\la\,\c_{\la'}^*$ lies in $\Omega_W(g,k)$
iff both $$\eqalignno{\bigl(T^W(g,k)\bigr)^{\dag}\,N^W\,\bigl(T^W(g,k)\bigr)=&
N^W,&(2.7a)\cr
\bigl(S^W(g,k)\bigr)^{\dag}\,N^W\,\bigl(S^W(g,k)\bigr)=&
N^W;&(2.7b)\cr}$$
$Z^{th}=\sum N^{th}_{\la\la'}\,t_\la \,t_{\la'}^*$ lies in $\Omega_{th}(g,k)$
iff both $$\eqalignno{\bigl(T^{th}(g,k)\bigr)^{\dag}\,N^{th}\,\bigl(T^{th}(g,k)
\bigr)=&N^{th},&(2.7c)\cr
\bigl(S^{th}(g,k)\bigr)^{\dag}\,N^{th}\,\bigl(S^{th}(g,k)\bigr)=&
N^{th}.&(2.7d)\cr}$$

The extension of these remarks to the semi-simple case is trivial.
 By a {\it type} $\T$ we mean the collection of ordered pairs
$$\T=\bigl(\{g_{1},k_{1}\},\{g_{2},k_{2}\},\ldots,\{g_{m},k_{m}\}\bigr),
\eqno(2.8a)$$
where each $g_i$ is a simple finite-dimensional Lie algebra, and $k_i$ is a
positive integer. Define
$$\eqalignno{P_{+}(\T)=&P_{+}(g_1,k_1)\times \cdots \times P_{+}(g_m,k_m),
&(2.8b)\cr
\l(\T)=&M_{g_1}^{(\h_1+k_1)}\oplus \cdots \oplus M_{g_m}^{(\h_m+k_m)},
&(2.8c)\cr
\c_\la(\T)(z,\u)=&\c_{\la_1}^{g_1,k_1}(z_1,\u)\cdots
\c_{\la_m}^{g_m,k_m}(z_m,\u),&(2.8d)\cr
t_\la(\T)(z,\u)=&t_{\la_1}^{g_1,k_1}(z_1,\u)\cdots
t_{\la_m}^{g_m,k_m}(z_m,\u),&(2.8e)\cr
W(\T)=&W(g_1)\times\cdots\times W(g_m),&(2.8f)\cr}$$
where `$\times$' in (2.8$b,f)$ denotes the cartesian product of sets and
`$\oplus$' in (2.8$c$) denotes the orthogonal direct sum of lattices, and where
in (2.8$d$) $\la=(\la_1,\ldots,\la_m)\in P_{+}(\T)$, in
(2.8$e$) $\la=(\la_1,\ldots,\la_m)\in \l(\T)^*/\l(\T)$, and in (2.8$d,e$)
 $z=(z_1,\ldots,z_m)\in {\bf C}\otimes \l(\T)$.

The modular matrices $S$ and $T$ become the matrix tensor products
$$\eqalignno{T^W(\T)=&T^W(g_1,k_1)\otimes\cdots\otimes T^W(g_m,k_m),&(2.9a)\cr
S^W(\T)=&S^W(g_1,k_1)\otimes\cdots\otimes S^W(g_m,k_m),&(2.9b)\cr
T^{th}(\T)=&T^{th}(g_1,k_1)\otimes\cdots\otimes T^{th}(g_m,k_m),&(2.9c)\cr
S^{th}(\T)=&S^{th}(g_1,k_1)\otimes\cdots\otimes S^{th}(g_m,k_m).&(2.9d)\cr}$$
The commutants of type $\T$ are written $\Omega_W(\T)$ and $\Omega_{th}(\T)$;
of course the analogues of eqs.(2.7) remain valid.

Clearly, if $Z_i\in \Omega_W(g_i,k_i)$ has coefficient matrix $N_i$, then
the function corresponding to $N_1\otimes \cdots \otimes N_m$ lies in
$\Omega_W(\T)$. The converse is not true, nor do such tensor products even
{\it span} $\Omega_W(\T)$, in general. A trivial example is
 $g_1=g_2=A_1$, $k_1=k_2=1$: each
$\Omega_W(g_i,k_i)$ here is only 1-dimensional, while $\Omega_W(\T)$ is
2-dimensional.

When $m=1$ in (2.8$a$) we call $\T$ a {\it simple type}; otherwise $\T$
is called {\it semi-simple}.
The classification of physical invariants of semi-simple type unfortunately
does not seem to reduce in any convenient way to the classification of
physical invariants of simple type.

\bigskip\bigskip \centerline{{\bf 3. The theta-commutant as a vector
space}}\bigskip

Our goal in this section is to extend the work of [14]. In particular, they
generated the $A_n$ commutant using certain orbits of SL(2,$\Z$). We will
extend their analysis, which was done only for $A_n$, to any semi-simple
 algebra. This will set the stage for the following section where we prove
the lattice partition functions of the Roberts-Terao-Warner method span
the commutant.

It should be noted that [14] is only
concerned with the structure of the commutant as a vector space.
The commutant has a much richer structure than that (\eg it is an algebra),
 and some of this
 additional structure will be exploited in Sec.5.

We will be concerned here solely with the theta-commutant. Because of that,
some of the labels `$th$' will be dropped.

Because we will be referring to these so frequently, call
$$G=\l(\T)^*/\l(\T),\sp\sp G_2=\bigl(\l(\T)\bigr)^{(2)*}/\bigl(\l(\T)
\bigr)^{(2)},$$ where as usual the superscript `$(2)$' refers to scaling
the lattice by $\sqrt{2}$. For any $\mu\in G_2$, by `$\sqrt{2}\mu$' we will
always mean the coset $\sqrt{2}\mu+\l(\T)\in G$.

To find a convenient description of $\Omega_{th}$, it will be necessary to
simplify the action of the modular matrices $S$ and $T$. To do this,
for each pair $\mu,\mu'\in G_2$ define a matrix $\{\mu,\mu'\}$ by
$$\{\mu,\mu'\}_{\la\la'}=\delta_{\la,\sqrt{2}\mu+\la'} \exp[2\pi i(\mu\cdot
\mu'+\sqrt{2}\mu'\cdot \la')],\eqno(3.1a)$$
for all $\la,\la'\in G$. This $|\l(\T)|\times |\l(\T)|$ matrix is the
coefficient matrix of
$$\sum_{\la,\la'}\{\mu,\mu'\}_{\la,\la'}\,t_\la(\T)\,t_{\la'}(\T)^*=
e^{-2\pi i\mu\cdot \mu'}\,{A^{(\sqrt{2}\mu;0),(\sqrt{2}\mu';0)}(\l_D
)\over D(\T)\,D(\T)^*},\eqno(3.1b)$$
where the function $A^{-,-}(\l)$, introduced in [6], is defined by
$$\eqalignno{A^{u,v}(\l)(z_Lz_R|\u)=&\sum_{(x_L;x_R)\in\l}\exp[\pi i \u
(x_L+u_L)^2-\pi i\u^*(x_R+u_R)^2]\cdot &(3.1c)\cr
&\cdot\exp[2\pi i\{(z_L+v_L)\cdot(x_L+u_L)-(z_R^*+v_R)\cdot(x_R+u_R)\}],&\cr}$$
 and where
$\l_D$ is the {\it diagonal gluing} for $\l(\T)$, \ie the even self-dual
lattice
$$\l_D=\l_D(\T)\equi\bigcup_{\la\in G} (\la;\la).\eqno(3.2)$$
These are easily shown to span all $|\l(\T)|\times |\l(\T)|$ complex matrices
$M_{\la,\la'}$ (see eqs.(3.7$b,c$) below), but they are not linearly
independent (see (3.7$d$) below). Their
value as linear generators of the matrices rests with these relations:
$$\eqalignno{T^{\dag}\,\{\mu,\mu'\}\,T=&\{\mu,\mu'+\mu\}\equi \{\mu,\mu'\}
\left( \matrix{1&1\cr 0&1\cr}\right), &(3.3a)\cr
S^{\dag}\,\{\mu,\mu'\}\,S=&\{\mu',-\mu\}\equi \{\mu,\mu'\}
\left( \matrix{0&-1\cr 1&0\cr}\right). &(3.3b)\cr}$$
The derivation of (3.3$a$) is straightforward, while (3.3$b$) follows most
easily from the transformation properties of $A^{-,-}(\l)$ under $\u\rightarrow
-1/\u$ (see eq.(3.8$b)$ in [6]).
As we know [16], the matrices $\left( \matrix{1&1\cr 0&1\cr}\right)$ and
$\left( \matrix{0&-1\cr 1&0\cr}\right)$ generate $\Gamma$=SL(2,$\Z$). This
immediately
suggests a description of $\Omega_{th}$. In particular, let $N=|\l(\T)^{(2)}|
=\|G_2\|$.
Then for all $\mu,\mu'\in G_2$,
$$\{\mu,\mu'\}\left( \matrix{a&b\cr c&d\cr}\right)\equi \{a\mu+c\mu',
b\mu+d\mu'\}\eqno(3.4a)$$
depends only on the values of $a,b,c,d$ (mod $N$), so the matrix
$$\sum_{K\in \Gamma_{2N}}\{\mu,\mu'\}K,\eqno(3.4b)$$
where $\Gamma_{N}=$SL(2,$\Z_{N})$,
commutes with $S$ and $T$ and lies in $\Omega_{th}(\T)$. In fact, because
the matrices in (3.1$a$) span all complex matrices, we immediately get that
 the matrices in (3.4$b$) span $\Omega_{th}(\T)$. Note that if we let
$\v$ be the $\Gamma$-orbit in $G_2\times G_2$ of $(\mu,\mu')$ --- \ie
the set $\v=\{(\mu,\mu')K\,|\,K\in \Gamma\}$ --- then ($3.4b$)
is an integral multiple of
$$N_\v\equi \sum_{(\mu_1,\mu_2)\in\v}\{\mu_1,\mu_2\},\eqno(3.4c)$$
and so these $N_\v$ span $\Omega_{th}(\T)$.

That observation is all that will be required for the following section, and
the proof that lattice partition functions span the commutant.
However, we will continue to follow [14] for now and explicitly find a
basis for $\Omega_{th}$. Our goal for the remainder of this section will
be to generalize their result for $\T=\bigl(\{A_n,k\}\bigr)$ (Thm.2 in [14]) to
any type $\T$.

Let $\v_1,\v_2$ be two $\Gamma$-orbits in $G_2\times G_2$, and suppose
some pairs $(\mu_i,\mu_i')\in \v_i$ satisfy $\sqrt{2}(\mu_1,\mu_1')=
\sqrt{2}(\mu_2,\mu_2')$, modulo $\l(\T)\times \l(\T)$ of course. Choose
any $K=\left( \matrix{a&b\cr c&d\cr}\right)\in\Gamma$. Because $\l(\T)$
is an {\it even} lattice, norms of $\la\in G$ are well-defined (mod 2). Hence,
$\mu_1^2\equiv\mu_2^2,\mu_1'{}^2\equiv\mu_2'{}^2,2\mu_1\cdot\mu'_1\equiv
2\mu_2\cdot\mu_2'$ (mod 1). That means that (using $ad-bc=1$)
$$\eqalignno{(a\mu_1+c\mu_1')\cdot(b\mu_1+d\mu_1')-\mu_1\cdot\mu_1'\equiv &
 ab\mu_1^2+cd\mu_1'{}^2+2bc\mu_1\cdot\mu_1' &(3.5a)\cr
\equiv& (a\mu_2+c\mu_2')\cdot(b\mu_2+d\mu_2')-\mu_2\cdot\mu_2'\sp({\rm mod}
\sp 1),&\cr}$$
which, together with (3.7$d$) below, gives us the matrix equality
$$\exp[2\pi i\mu_1\cdot \mu_1']\,N_{\v_1}=r\cdot\exp[2\pi i\mu_2\cdot \mu_2']
\,N_{\v_2},\eqno(3.5b)$$
where $r$ is a positive rational number calculated from the number  of times
$\sqrt{2}\v_i$ covers the orbits $\Gamma(\sqrt{2}\mu_i,\sqrt{2}\mu_i')$,
$i=1,2$.
Thus the invariants corresponding to $\v_1$ and $\v_2$ differ only
by a constant factor.

Thus to any $\Gamma$-orbit $\tilde{\v}$ in $G\times G$ (as opposed to
$G_2\times G_2$), we can assign an
invariant $M_{\tilde{\v}}$ in $\Omega_{th}(\T)$, well-defined up to a constant
phase, defined in the following manner: let $\v$ be any $\Gamma$-orbit
in $G_2\times G_2$ for which $\sqrt{2}\v=\tilde{\v}$, and define
$$M_{\tilde{\v}}\equi N_{\v}.\eqno(3.6)$$
Obviously, many such orbits $\v\subset G_2\times G_2$ exist; (3.5$b$) shows
that which orbit $\v$ is chosen will only affect the answer $M_{\tilde{\v}}$
by a constant (hence irrelevant) phase factor.

\bigskip\noindent{\bf Theorem 1}:\quad (a) For any semi-simple $\T$, let
${\cal C}(\T)$ be the set of all $\Gamma$-orbits $\tilde{\v}$ of $G\times G$
for which $M_{\tilde{\v}}$ is not the zero matrix. Then the set
$$\{M_{\tilde{\v}}\,|\,\tilde{\v}\in{\cal C}(\T)\}$$
is a basis for $\Omega_{th}$.

\item{(b)} For $g=A_{2\ell}$, $B_{4\ell}$, $D_{4\ell}$, $E_6$, $E_8$, $F_4$
and $G_2$, ${\cal C}$ is the set of all $\Gamma$-orbits of $G\times G$.
For $g=C_n,$ $E_7$, and the remaining $B_n$ and $D_n$, ${\cal C}$ consists
of those orbits $\tilde{\v}$ satisfying:
$$(\la,\la')\in\tilde{\v} \Rightarrow (k+\h)\la^2\equiv (k+\h)\la'{}^2\equiv 0
 \sp({\rm mod}\sp 1).$$
For $g=A_{2\ell-1}$, the condition on $\tilde{\v}$ is:
$$(\la,\la')\in\tilde{\v} \Rightarrow (k+\h)\la^2\equiv (k+\h)\la'{}^2\equiv 0
 \sp({\rm mod}\sp {2\over \ell}).$$\bigskip

[14] proved this theorem for $g=A_n$, and used it to calculate the dimension
of $\Omega_{th}(A_2,k)$ (and hence $\Omega_{th}(G_2,k-1)$). [17] later used
 the theorem to compute the
dimension for all $A_n$. The calculation was sufficiently general that,
together with the above theorem, it should now be possible to extend this
dimension calculation to any simple type. However, the formulae obtained
in [17] for $n>2$ was sufficiently complicated that the value of extending
this work to all other simple types is questionable.

\bigskip
\noindent{\it Proof}\quad First define, for each $\la,\la'\in G$, the matrix
$\widetilde{\{\la,\la'\}}$ by
$$\widetilde{\{\la,\la'\}}_{\la_L,\la_R}\equi \delta_{\la_L,\la+\la_R}\,
\exp[2\pi i(\la\cdot\la'+\la'\cdot \la_R)],\eqno(3.7a)$$
where $\la_L,\la_R\in G$. It is easy to show that these span all $|\l(\T)|
\times |\l(\T)|$ matrices: for any $\la_1,\la_2\in G$ let $E(\la_1,\la_2)$
denote the matrix
$$E(\la_1,\la_2)_{\la_L\la_R}=\delta_{\la_1\la_L}\cdot\delta_{\la_2\la_R}$$
consisting of zeros everywhere except for one `1' at $(\la_1,\la_2)$. These
$E(\la_1,\la_2)$ span all matrices. The usual projection argument gives
$$E(\la_1,\la_2)={1\over \|G\|}\sum_{\la\in G} e^{-2\pi i\la\cdot \la_1}
\widetilde{\{\la_1-\la_2,\la\}}.\eqno(3.7b)$$
Therefore the $\widetilde{\{\la,\la'\}}$ must also span all matrices.
Furthermore, a dimension check tells us they constitute a {\it basis}.

Incidently, note that for any $\mu,\mu'\in G_2$,
$$\{\mu,\mu'\}=e^{-2\pi i\mu\cdot\mu'}
\widetilde{\{\sqrt{2}\mu,\sqrt{2}\mu'\}},
\eqno(3.7c)$$
so if $\sqrt{2}\mu_1\equiv \sqrt{2}\mu_2,\sqrt{2}\mu_1'\equiv \sqrt{2}\mu_2'$
(mod $\l(\T)$), then
$$\{\mu_2,\mu_2'\}=e^{2\pi i(\mu_1\cdot\mu_1'-\mu_2\cdot \mu_2')}\{\mu_1,\mu_1'
\}.\eqno(3.7d)$$
Because the $\widetilde{\{\la,\la'\}}$ are linearly independent, eq.(3.7$d$)
generates all linear relations among the $\{\cdot,\cdot\}$.

(a) now follows: different
orbits $\tilde{\v}$ of $G\times G$ correspond to $M_{\tilde{\v}}$
being linear combinations
of {\it disjoint} sets of terms $\widetilde{\{\la,\la'\}}$, and since these
terms are all linearly independent, so must the nonzero $M_{\tilde{\v}}$. We
already know they span $\Omega_{th}$, and hence they form a basis.

To prove (b), we will first make some general calculations, and then
conclude the proof by looking explicitly at $g=C_n$. The other algebras can
be done similarly (in most cases, somewhat more easily, too).

Choose any $\Gamma$-orbit $\v$ in $G_2\times G_2$. We want to find what
conditions $\sqrt{2}\v$ must satisfy for $N_\v$ to be non-zero. For any
$\la,\la'\in G$, define $[\la,\la']$ to be the (possibly empty) set of all
pairs
$(\mu,\mu')\in \v$ for which $(\sqrt{2}\mu,\sqrt{2}\mu')=(\la,\la')$.
Then
$$N_\v=\sum_{\la,\la'\in G}\widetilde{\{\la,\la'\}}
\sum_{(\mu,\mu')\in[\la,\la']}
\exp[-2\pi i \mu\cdot \mu']\equi \sum_{\la,\la'\in G}s_{\la\la'}
\widetilde{\{\la,\la'\}}.\eqno(3.8)$$
$N_\v\not= 0$ iff some $s_{\la\la'}\neq 0$, which turns out to happen iff
for some $(\la,\la')\in
\sqrt{2}\v$, $\mu\cdot \mu'$ is constant (mod 1) for all $(\mu,\mu')\in [\la,
\la']$.

Those comments hold for any (simple or semi-simple) $\T$. Now, turn to $g=C_n$.
Let $\ell=k+\h$. $M^{(\ell)}$ here is the orthogonal lattice $A_1^{(\ell)}
\oplus \cdots\oplus A_1^{(\ell)}$. Let $e_i$ be an orthonormal basis, then
$G_2$ is generated by the cosets
$$g_i+M^{(\ell)},\sp\sp{\rm where}\sp g_i={e_i\over \sqrt{4\ell}}.$$ Each of
these cosets has order $4\ell$.

Now let $(\mu,\mu'),(\mu,\mu')K$ both lie in $[\la,\la'],$ where
$K=\left(\matrix{a&b\cr c&d\cr}\right)$. We may write $\mu=\sum x_i g_i$,
$\mu'=\sum x_i'g_i$ for $x_i,x_i'\in \Z_{4\ell}$. Then we have
$$ax_i+cx_i'\equiv x_i,\sp\sp bx_i+dx_i'\equiv x_i'\sp ({\rm mod}\sp 2\ell).
\eqno(3.9a)$$
We are interested in computing $$(a\mu+c\mu')\cdot (b\mu+d\mu')-\mu\cdot\mu'
={1\over 4\ell}\sum_{i=1}^n[(ax_i+cx_i')\cdot(bx_i+dx_i')-x_ix_i'],
\eqno(3.9b)$$
because we know that $s_{\la\la'}$ will be nonzero iff all $K\in\Gamma$
satisfying $(3.9a)$ necessarily have $(3.9b)$ congruent to 0 (mod 1).

Suppose an odd number of $x_i$ are odd. Then consider $K=\left( \matrix{
1&2\ell\cr 0&1\cr}\right)$. Clearly $K$ satisfies $(3.9a)$. Eq.$(3.9b)$ becomes
$${1\over 4\ell}\sum_{i=1}^n x_i\cdot 2\ell x_i={1\over 2}\sum_{i=1}^n
x_i^2 \equiv {1\over 2}\sp({\rm mod\sp} 1),$$
and so $s_{\la\la'}$ would vanish. A similar calculation holds when an
odd number of $x_i'$ are odd.

Now suppose an even number of both $x_i$ and $x_i'$ are odd. We wish to
show $s_{\la\la'}\neq 0$ in this case. Then $\mu,\mu'\in 1/\sqrt{4\ell}D_n$,
so we can express these in terms of the simple root basis $\alpha_i$ of
$D_n$ (enumerated as in Table Fin of [3]):
$$\mu=\sum_{i=1}^{n-1}y_i {\alpha_i \over \sqrt{4\ell}}+y_n{2e_n\over
\sqrt{4\ell}},\sp\sp
\mu'=\sum_{i=1}^{n-1}y'_i {\alpha_i \over \sqrt{4\ell}}+y'_n{2e_n\over
\sqrt{4\ell}}\eqno(3.10a)$$
(we substituted $2e_n$ for $\alpha_n$ to simplify eqs.($3.10b,c)$ below).
Eq.($3.9a)$ becomes
$$\eqalignno{ay_i+cy_i'\equiv y_i,\sp\sp & by_i+dy_i'\equiv y_i'\sp ({\rm mod}
\sp 2\ell),\sp \sp i=1,\ldots,n-1,&(3.10b)\cr
ay_n+cy_n'\equiv y_n,\sp\sp &by_n+dy_n'\equiv y_n'\sp ({\rm mod}\sp \ell).
&(3.10c)\cr}$$
Eq.(3.9$b$) now becomes
$$\eqalignno{{2\over 4\ell}\sum_{i=1}^{n-1} [(ay_i+cy_i')(by_i+dy_i')-
y_iy_i']&
+{4\over 4\ell}[(ay_n+cy_n')(by_n+dy_n')-y_ny_n']&(3.10d)\cr
-{1\over 4\ell}\sum_{i=1}^{n-2} [(ay_i+cy_i')(by_{i+1}+dy_{i+1}')-
y_iy_{i+1}']&
-{1\over 4\ell}\sum_{i=1}^{n-2} [(ay_{i+1}+cy_{i+1}')(by_i+dy_i')-
y_{i+1}y_i']&\cr
-{2\over 4\ell}[(ay_{n-1}+cy_{n-1}')(by_n+dy_n')-y_{n-1}y_n']&
+{4\over 4\ell}[(ay_n+cy_n')(by_{n-1}+dy_{n-1}')-y_ny_{n-1}'].&
}$$
The following argument shows this must necessarily be $\equiv 0$ (mod 1). Each
$(ay_i+cy_i')\cdot(by_i+dy_i')-
y_iy_i'\equiv 0$ (mod 2$\ell$) by $(3.10b)$, so (mod 1) the first term
vanishes.
Similarly, $(3.10c$) implies the second term vanishes (mod 1). $(3.10b)$
tells us each
$(ay_{i+1}+cy_{i+1}')\cdot(by_i+dy_i')-
y_{i+1}y_i'\equiv 0$ (mod $2\ell$), so (mod 1) we may replace the $-{1\over
4\ell}$ coefficient of the fourth term with $+{1\over 4\ell}$, with the
result (using $ad-bc=1$) that the third term exactly cancels with this modified
fourth term. A similar calculation shows that the final two
terms cancel (mod 1).

The condition that there be an even number of odd $x_i$ and $x_i'$ is clearly
equivalent to the condition that $\la^2\equiv \la'{}^2\equiv 0$ (mod
$1/\ell)$),
and if it holds for one pair $(\la,\la')\in \sqrt{2}\v$, it holds for all
pairs. \qquad QED\bigskip

[14] went on to prove that $\Omega_{th}$ had an {\it integral basis}; we
will do this in the following section by showing that lattice partition
functions span the commutant.

\bigskip\bigskip\centerline{{\bf 4. Lattice partition functions
and the commutant}}\bigskip

For some purposes (\eg calculating dimensions), knowing an explicit basis
can be helpful. However, the matrix elements of the $N_\v$ or $M_{\tilde{\v}}$
defined in the previous section will in general be complex,
and in any event those matrices are far from
conducive to practical calculations, so are of limited value in the search
for physical invariants.

A considerably more practical means of generating invariants in $\Omega_{th}$
or $\Omega_W$ is the lattice method of Roberts-Terao-Warner. We will very
briefly review it below. A more thorough presentation is provided in [6,12,13].

An {\it integral} lattice is one in which all dot products are integers;
it is called  {\it even} if in addition all its norms are even.
 A {\it self-dual} lattice $\l$ is one which equals
 its dual $\l^*$. $\l$ is self-dual iff it is integral and also has determinant
$|\l|=1.$ Finally, by a {\it gluing} $\l$ of $\l_0$, we mean that $\l_0$
is a sublattice in $\l$, and $\l/\l_0$ is a finite group.
By $(\l_0;\l_0)$ we mean the indefinite lattice with elements $x=(x_L;x_R)$,
$x_L,x_R\in \l_0$, whose dot products are defined by $x\cdot x'=x_L\cdot x_L'
-x_R\cdot x_R'$.

Consider any lattice $\l$. Define its type $\T$ {\it partition function}
$$Z_\l(\T)(z_L,z_R|\u) \equi {
 \sum_{(x_L;x_R)\in\l} \exp[\pi i \u x_L^2-\pi i \u^* x_R^2+2\pi i(z_L'
\cdot x_L-z_R'{}^*\cdot x_R)]\over D\bigl(\T\bigr)(z_L|\u)\cdot D\bigl(\T\bigr)
(z_R|\u)^*},\eqno(4.1a)$$
where we use $z'=(\sqrt{k_1+\h_1}z_1,\ldots,\sqrt{k_m+\h_m}z_m)$.
If $\l$ is a gluing of $\bigl(\l(\T);\l(\T)\bigr)$, we may write this as
(again using $G=\l(\T)^*/\l(\T)$)
$$\eqalignno{Z_\l\bigl(\T\bigr)(z_L,z_R|\u) \equi &
  \sum_{\la_L,\la_R\in G} \bigl(N_\l\bigr)_{\la_L\la_R}\,
t_{\la_L}(\T)(z_L|\u)\,\,t_{\la_R}(\T)(z_R|\u)^*&(4.1b)\cr
{\rm where}\sp\sp&\bigl(N_\l\bigr)_{\la_L\la_R}=\left\{ \matrix{1&{\rm if}
\sp(\la_L;\la_R)\subset \l\cr 0&{\rm otherwise}} \right..&(4.1c)\cr }$$
We will call $N_\l$ the coefficient matrix of $Z_\l(\T)$ or, more briefly,
the coefficient matrix of $\l$.

If $\l$ is in addition both even and self-dual, $Z_\l(\T)$ will be modular
invariant. Thus, finding all even self-dual gluings $\l$ of $\bigl(\l(\T);
\l(\T)\bigr)$,
and computing each of their partition functions $Z_\l(\T)$, constitutes a
method of generating elements in $\Omega_{th}$. Weyl-folding these (see
eqs.(2.3),(2.5)) produces functions $WZ_\l(\T)\in \Omega_W(\T)$:
$$WZ_\l\bigl(\T\bigr)(z_L,z_R|\u) \equi \sum_{w,w'}\epsilon(w)\epsilon(w')
  Z\bigl(\T\bigr)(w(z_L),w'(z_R)|\u),\eqno(4.1d)$$
where the sum is over all $w,w'$ in the Weyl group $W(\T)$ (that $WZ_\l$
can be written as a linear combination of $\c_\la(\T)\,\c_{\la'}(\T)^*$
for $\la,\la'\in P_{+}(\T)$ follows from eqs.(2.5)).
 We will call this approach the
{\it Roberts-Terao-Warner} method.

Let ${\cal L}(\T)$ denote the set of all even self-dual gluings of
$\bigl(\l(\T);\l(\T)\bigr)$. It always is finite. Note that it is
always non-empty: see (3.2).
In fact, $\l_D$
corresponds to $N_{\l_D}=I$, the identity matrix. The partition functions
$Z_{\l_D}$ and $WZ_{\l_D}$ are often called the {\it diagonal invariants}.

Let $\Omega^L_{th}(\T)$ be the (complex) space spanned by all $Z_\l(\T)$,
and let $\Omega_W^L(\T)$ be the (complex) space spanned by all $WZ_\l(\T)$,
for $\l\in{\cal L}(\T)$ --- they will be called the {\it lattice
theta-commutant} and {\it lattice Weyl-folded commutant}, respectively.
The previous discussion tells us that $\Omega^L_{th}(\T)\subseteq \Omega_{th}
(\T)$ and $\Omega^L_W(\T)\subseteq \Omega_W(\T)$.

\bigskip\noindent{\bf Theorem 2}:\quad (a) The lattice theta-commutant
equals the theta-commutant for any type $\T$:
$$\Omega^L_{th}(\T)= \Omega_{th}(\T);\eqno(4.3a)$$

\item{(b)} the lattice Weyl-folded commutant equals the Weyl-folded
commutant, for any $\T$:
$$\Omega^L_W(\T)= \Omega_W(\T).\eqno(4.3b)$$\bigskip

Hence the Roberts-Terao-Warner lattice method is complete. The $Z_\l$ are
far from linearly independent, and following this theorem
we will describe a small class of lattices for which the $Z_\l$ still span the
commutant.

The Weyl-Kac formula tells us that any $Z\in \Omega_{W}$ can be written
as the sum of partition functions of translates of the lattice $\bigl(\l(\T);
\l(\T)\bigr)$.
However, it does not follow immediately from modular invariance that those
translates can be grouped together in such a way that $Z$ can be written
as a sum of $WZ_\l$ for $\l\in {\cal L}(\T)$. In other words, the direct
approach to proving (4.3$b$) does not appear promising.

This theorem proves that $\Omega_{th}(\T)$ and  $\Omega_W(\T)$ always have
integral bases. We are more interested in $(4.3b)$, but it is a trivial
corollary of $(4.3a)$, the equality we will prove. We will go about this
by defining a set of lattices $\l(\mu\mu';a)\in {\cal L}(\T)$ in (4.4),
expressing in (4.7) their coefficient matrices $N_{\l(\mu\mu';a)}$ (defined
as in (4.1$c$)) in terms of the $N_\v$ of $(3.4c)$, and then inverting those
relations so as to express the $N_\v$ in terms of the $N_{\l(\mu\mu';a)}$.
Since we know the $N_\v$ span $\Omega_{th}$, this would imply that the
$N_{\l(\mu\mu';a)}$ span $\Omega_{th}$, and hence that $\Omega^L_{th}=
\Omega_{th}$.

\bigskip \noindent{\it Proof}\quad Consider any $\Gamma$-orbit $\v$ in
$G_2\times G_2$. We want to show $N_\v\in \Omega^L_{th}$. For any
$(\mu,\mu')\in \v$ we can define the lattice $L(\v)$ spanned (over
the integers) by $\mu$, $\mu'$ and $\l(\T)^{(2)}$; this lattice is independent
of which pair in $\v$ is chosen. $L(\v)$ will in general be neither
self-dual nor integral.

Call the {\it size} of $\v$ the number $s(\v)\equi |\l(\T)^{(2)}|/|L(\v)|$;
it will always be an integer, and in fact a perfect square.

Our proof will be by induction on the size of $\v$.

For $\v$ with size $s(\v)=1$, $L(\v)=\l(\T)^{(2)}$ and $\v=\{(0,0)\}$.
 In this
case, $N_\v=\{0,0\}=I$, the identity matrix, which we know equals
$N_{\l_D}\in\Omega^L_{th}$.

Now consider any $\Gamma$-orbit $\v$ in $G_2\times G_2$, and assume that
$N_{\v'}\in \Omega^L_{th}$ for any orbit $\v'$ with $s(\v')<s(\v)$.

$L(\v)/\l(\T)^{(2)}$ will be an abelian group isomorphic to
$\Z_d\times\Z_{dd'}$, for some $d,d'$. (Incidently, $s(\v)=d^4d'{}^2$.)

Choose any $(\mu,\mu')\in \v$, and for $a=0,1,\ldots,d-1$ define the
lattice $\l(\mu,\mu';a)$ to be the {\it shifting}
$$\eqalignno{\l(\mu,\mu';a)\equi & \l_D\bigl(\{(\sqrt{2}\mu;0),
(\sqrt{2}\mu';0)\},\zeta\bigr),&(4.4a)\cr
{\rm where}\sp & \zeta=\left( \matrix{-\mu^2&-\mu\cdot \mu'-{a\over d}\cr
-\mu\cdot \mu'+{a\over d}&-\mu'{}^2\cr}\right);&(4.4b)\cr}$$
\ie $\l(\mu,\mu';a)$ is the lattice given by
$$\eqalignno{\l(\mu,\mu';a)=\{&(\la+\ell \sqrt{2}\mu+\ell'\sqrt{2}\mu';\la)
+\bigl(\l(\T);\l(\T)\bigr)\,|\,\ell,\ell'\in\Z,
\sp \la\in\l(\T)^*,\sp &\cr &{\rm and}\sp \sqrt{2}\la\cdot \mu\equiv
\ell\zeta_{11}
+\ell'\zeta_{12}, \sp \sqrt{2}\la\cdot \mu'\equiv
\ell\zeta_{21}+\ell'\zeta_{22}\sp({\rm mod}\sp 1)\,\}.&(4.4c)\cr}$$

Shifting is discussed in reasonable detail and generality in [18].
For now it suffices to remark that $\l(\mu,\mu';a)$ will be self-dual (this can
also be read off from (4.7) below, as was done in the claim in Thm.E of [6]).
It is clear from its definition that
it is also even, and is a gluing of $\bigl(\l(\T);\l(\T)\bigr)$.

\bigskip \noindent{\bf Claim:}\quad $\l(\mu,\mu';a)$ has the coefficient matrix
(see (4.1$c$))
$$N_{\l(\mu,\mu';a)}(\T)=
c\sum\exp[2\pi i
{a\over d}(-\ell_1\ell_2'+\ell_2\ell_1')] \, \{\gamma_1,\gamma_2\},\eqno(4.5)$$
where the sum is over all $\gamma_1=\ell_1\mu+\ell_1'\mu',\gamma_2=\ell_2
\mu+\ell_2'\mu'\in L(\v)$, and where $c=1/\sqrt{s(\v)}$. \bigskip

The proof involves using (3.1$a$) to rewrite the RHS of (4.5), showing that it
equals the matrix $N_{\l(\mu,\mu';a)}$
corresponding to eq.(4.4$c$). The key point in this straightforward
derivation is the observation that for any $\la\in G$,
$$\sum_{\gamma_2}\exp[2\pi i \{
{a\over d}(-\ell_1\ell_2'+\ell_2\ell_1')+\sqrt{2}\gamma_2\cdot \la+\gamma_1
\cdot\gamma_2\}]\eqno(4.6a)$$
vanishes iff there exists a $\gamma_3=\ell_3\mu+\ell_3'\mu'\in L(\v)$ such that
$$ {a\over d}(-\ell_1\ell_3'+\ell_3\ell_1')+\sqrt{2}\gamma_3\cdot \la+\gamma_1
\cdot\gamma_3\not\equiv 0\sp({\rm mod}\sp 1).\eqno(4.6b)$$
The expression in (4.6$b$) will be $\equiv 0$ (mod 1) for all $\gamma_3$, iff
it is $\equiv 0$ (mod 1) for $\gamma_3=\mu$ and $\gamma_3=\mu'$ --- these are
precisely the two congruences in (4.4$c$).

We can rewrite $(4.5)$ in a more convenient form, using the easily verified
fact (Lemma 2(i) of [14]) that, for any $\Gamma$-orbit $\v'$ with $L(\v')
\subseteq L(\v)$, $B(\v')\equi\ell_1\ell_2'-\ell_2\ell_1'$ is
 independent
 (mod $d$) of which pair $(\ell_1\mu+\ell_1'\mu',\ell_2\mu+\ell_2'
\mu')$ is chosen from $\v'$.
The result is:
$$N_{\l(\mu,\mu';a)}=c\sum_{\v'}\exp[-2\pi i B(\v'){a\over d}]\,N_{\v'}
\eqno(4.7)$$
where the sum is over all $\Gamma$-orbits $\v'$ whose lattices $L(\v')$ are
sublattices of $L(\v)$.

It is important to realize that there is precisely one orbit $\v'$
with $L(\v')=L(\v)$, for each $0<m\le d$ relatively prime to $d$
(the assignment is given by $m=B(\v')$) (see p.622 of
[14]). Of course $\v'=\v$ is the unique one corresponding to $m=B(\v')=1.$

By the induction hypothesis we know $N_{\v''}\in \Omega^L_{th}$ for any
orbit $\v''$ in (4.7) with size $s(\v'')$ less than $s(\v)$. $N_\v$ can be
solved for in terms of these $N_{\v''}$ and the $N_{\l(\mu,\mu';a)}$,
by multiplying (4.7) by $\exp[2\pi i a/d]$ and summing over all $a$.

Hence $N_\v\in \Omega^L_{th}$.\qquad QED
\bigskip

The class of lattices in (4.4) shown to span $\Omega_{th}$ is small compared
to the whole class ${\cal L}$ of even self-dual gluings of $\bigl(\l(\T);
\l(\T)\bigr)$. We learn from the proof of Thm.2 that $\Omega_{th}$ is spanned
by the partition functions $Z_{\l(\mu,\mu';a)}(\T)$, where $\mu,\mu'\in G_2$.
Eq.(4.7) shows that this is a function of $\v$ and $a$ --- \ie if $(\mu_1,
\mu_1')$ and $(\mu_2,\mu_2')$ lie on the same $\Gamma$-orbit $\v$, then
$\l(\mu_1,\mu_1';a)=\l(\mu_2,\mu_2';a)$. A similar argument shows that if
$L(\v_1)=L(\v_2)$, then $\l(\mu_1,\mu_1';a)=\l(\mu_2,\mu_2';B_{\v_1}(\v')a)$.
Moreover, if $(\sqrt{2}\mu_1,\sqrt{2}\mu_1')=(\sqrt{2}\mu_2,\sqrt{2}\mu_2')$,
then $(3.5b)$ and (4.7) tell us that the $N_{\v_2'}$ in (4.7) can be recovered
from the $\l(\mu_1\mu_1';a)$.

Finally, note that there was an element of overkill in the proof of Thm.2.
In particular, there we had `$a$' range from 0 to $d-1$. Let $k_1,\ldots,k_n$
be the $n=\phi(d)$ integers between 0 and $d$ relatively prime to $d$. All
that was relevant for the `inverting' step in the final paragraph of the proof
was that the $d\times n$ matrix with entries
$$B_{aj}=\exp[-2\pi i a\cdot k_j],\eqno(4.8)$$
for $a=0,1,\ldots,d-1$, $j=1,\ldots,n$, be of  rank $n$. It was that fact
which allowed us to solve (4.7) for the $N_{\v'}$ in terms of the
$N_{\l(\mu\mu'
;a)}$. However, the $n\times n$ submatrix obtained by restricting (4.8) to
$a=0,\ldots,n-1$, can also be shown to be of rank $n$ (\ie invertible). For
otherwise it would have a zero eigenvalue, which would mean there would
exist $\alpha_1,\ldots,\alpha_n\in {\bf C}$, not all zero, for which
$$0=\sum_{j=1}^n \alpha_j \, x^{j-1} \eqno(4.9)$$
for $x=\exp[-2\pi i k_1],\ldots,\exp[-2\pi i k_n]$. In other words, the
$n-1$-th degree polynomial in (4.9) would have $n$ distinct roots. This is
impossible. Thus our $n\times n$ submatrix is invertible.

We can use these remarks to significantly reduce the number of lattices we
need to consider. The result is as follows.

Consider each (not necessarily integral) lattice $L$ of the form
$$L=\bigcup_{i,j\in \Z} i\la+j\la'+\l(\T)\equi \l(\T)[\la,\la']$$
for some $\la,\la'\in \l(\T)^*$. Put $\mu=\la/\sqrt{2},$ $\mu'=\la'/\sqrt{2}$,
and define $d_L$ by
$$L'/\l(\T)^{(2)}\cong \Z_{d_L}\times \Z_{d_Ld_L'},\,\,{\rm where}\,\,
L'=\bigcup_{i,j\in\Z} i\mu+j\mu'+\l(\T)^{(2)}.$$
Now construct the lattices $\l(L,a)\equi\l(\mu,\mu';a)$ defined in eqs.(4.4),
but now for $a=0,1,\ldots,\phi(d_L)-1$.

Finally, let ${\cal L}_*(\T)$ denote the resulting collection of all these
$\sum_L \phi(d_L)$ even self-dual lattices $\l(L,a)$.

\bigskip\noindent{\bf Corollary 2}: \quad The $Z_\l(\T)$ for $\l\in{\cal L}_*
(\T)$, span $\Omega_{th}$; the corresponding $WZ_\l(\T)$ span $\Omega_W(\T)$.
\bigskip

Particularly for algebras of large rank, the set ${\cal L}_*(\T)$ is
considerably smaller than the set ${\cal L}(\T)$ of all even self-dual gluings
of $\bigl(\l(\T);\l(\T)\bigr)$. Moreover, for a fixed choice of algebra, it
should be possible to use Thm.1 to reduce ${\cal L}_*$ further so that the
corresponding $Z_\l$ will constitute a {\it basis} for $\Omega_{th}$.

One final remark will be made here. Q.{} Ho-Kim [19] is currently running
a program designed to compute all lattice partition functions $WZ_\l$ for
$\l\in {\cal L}(g,k)$, for $g=A_2,C_2,G_2,A_1+A_1$ and with levels $k$ up to
around 30.
He will in this way determine $\Omega^L_W$ and
hence $\Omega_W$ for these types, and thus compute
all physical invariants. The analysis is not yet complete, but when it is
it will apparently constitute the only completeness proof for these $g$,
for small $k$, apart from $k=1$ (see [11], as well as Thm.5 below), and
$k+3$ prime with $g=A_2$ [20]. (Various other computer searches have been
undertaken before, \eg in [13], but they do not exhaust the commutant and
so leave open the possibility that an unknown physical invariant may have
escaped detection.) One thing apparent from the work in [19] is the
practicality of this lattice method, at least for small $g$ and $k$.

\bigskip\bigskip\centerline{{\bf 5. General observations, and applications
to level 1}}\bigskip

In this section we make a few general observations relevant to the
classification problem, and use some of these to complete the classification
of all level 1 physical invariants of simple type.

Using the finite-dimensional Weyl denominator formula, a simpler
expression can be found (see eq.(13.8.10) in [3]) for the $\la=0$ row and
column
of the $S=S^W(\T)$ matrix in (2.6$d$): if $\T=\bigl(\{g,k\}\bigr)$, then for
any
$\la\in P_{+}(g,k)$ we have
$$S_{\la,0}=S_{0,\la}={1\over\sqrt{|M|}\cdot (k+\h)^{n/2}} \prod_{\alpha
\in \Delta_+} 2\,{\rm sin}{\pi (\la +\rho)\cdot \alpha \over k+\h},
\eqno(5.1)$$
where $\Delta_+$ is the set of positive roots of $g$. The corresponding
formula for semi-simple types is obtained from (5.1) by multiplication (see
(2.9$b$)). Because each $0\le\la\cdot \alpha\le k$ and $0<\rho
\cdot \alpha <\h$, we immediately get that each $S_{\la,0}$ is strictly
positive. Also, $S_{\la,0}\ge S_{0,0}$ and $\sum S_{\la,0}^2=1$.

This positivity has a number of easy consequences. For one thing, {\it the
number of physical invariants of a given type} $\T$ {\it must be finite}. To
see this let $s=S^W(\T)_{00}=$min${}_\la \{S^W(\T)_{\la,0}\}$ and let
$N_{\la\la'}$ be the
 coefficient matrix of any physical invariant of type $\T$. Then
$$1=N_{00}=\sum_{\la,\la'} S_{\la,0}\,S_{\la',0}\,N_{\la\la'} \ge s^2
\sum_{\la,\la'}N_{\la\la'}. \eqno(5.2)$$
Since each $N_{\la\la'}$ must be a non-negative integer, which by (5.2) is
bounded above by $1/s^2$, the desired finiteness follows.
Another consequence of this positivity is that if $N$ is any {\it positive}
invariant (not necessarily physical), then a similar calculation to that
given in (5.2) shows $N_{00}>0$.

A second observation connects more directly with the lattice formalism of
the previous section. Consider any type $\T$, and let $\l_0$ denote the
indefinite lattice $\bigl(\l(\T);\l(\T)\bigr)$ defined in Sec.4. Choose any
 $x=(x_L;x_R)\in \l_0^*$ and let its order be $m$ (\ie $\ell x\in \l_0$ iff
$m$ divides $\ell$). Let $\l_1$ be any gluing of $\l_0$ (\ie the quotient group
$\l_1/\l_0$ exists and is finite). Then for any $\ell$
relatively prime to $m$, $\ell$ has an inverse (mod $m$), so $x\in \l_1$ iff
$\ell x\in \l_1$.

Now for any $y=(y_L;y_R)\in \l^*_1$ consider the function
$$c(y)(z_Lz_R|\u)\equi \sum \epsilon(w)\epsilon(w') \, t_{w(y_L)}(\T)(z_L,\u)
\,t_{w'(y_R)}(\T)(z_R,\u)^*, \eqno(5.3a)$$
where the sum is over all $w,w'$ in the Weyl group $W(\T)$. Using eqs.(2.5)
we get that either
$$c(y)(z_Lz_R|\u)=0\eqno(5.3b)$$
for all $z_L,z_R,\u$, or there exist unique $\epsilon_y\in \{\pm 1\}$,
$\la_y,\la_y'\in P_{+}(\T)$ such that
$$c(y)(z_Lz_R|\u)=\epsilon_y\,\c_{\la_y}(z_L,\u)\,\c_{\la_y'}(z_R,\u)^*
\eqno(5.3c)$$
for all $z_L,z_R,\u$. It can be shown that for $\ell$ relatively prime to
$m$, $c(x)=0$ iff $c(\ell x)=0$.

So what does all this tell us? Note that we may write
$$WZ_{\l_1}(\T)=\sum_{y\in\l_1/\l_0} c(y)=\sum_{\la,\la'\in P_{+}(\T)}
N_{\la\la'}\,\c_{\la}(\T)\,\c_{\la'}(\T)^*.\eqno(5.3d)$$
Then this argument implies for any $x\in \l_1/\l_0$ for which $c(x)\ne 0$,
$$\epsilon_x\,N_{\la_x\la_x'}=\epsilon_{\ell x}\, N_{\la_{\ell x}\la_{\ell
x}'}.
\eqno(5.3e)$$
By Thm.2, eq.(5.3$e$) also holds for the coefficient matrix $N$ of {\it any}
$Z\in \Omega_W(\T)$, and for {\it any} $x\in\l_0^*/\l_0$.

For example, for $g=A_2$, $k=5$, (5.3$e$) gives us relations such as
$$N_{00,00}=N_{22,22} \sp {\rm and} \sp N_{10,10}=N_{01,01}=N_{40,40}=N_{04,04}
,\eqno(5.4a)$$
where we write the subscript `$ij$' for the vector $i\beta_1+j\beta_2$ with
Dynkin weights $i$ and $j$.
For $g=A_2$, $k=9$, we get relations such as
$$\eqalignno{N_{01,05}&=-N_{16,04}=-N_{12,54}=N_{08,45}=N_{18,40}=-N_{62,50}&\cr
=N_{10,50}&=-N_{61,40}=-N_{21,45}=N_{80,54}=N_{81,04}=-N_{26,05}.&(5.4b)}$$
For example, (5.4$b$) comes from $x=({\beta_1+2\beta_2\over \sqrt{12}};
{\beta_1+6\beta_2\over \sqrt{12}})$, which has order $m=36$, and $\ell=1$, 5,
7,
 11, 13, 17, $-1$, $-5$, $-7$, $-11$, $-13$, $-17$, respectively.
Note that because of the sign changes in (5.4$b$), any {\it positive} (hence
any {\it physical}) invariant of type $\bigl(\{A_2,9\}\bigr)$ must have
$0=N_{01,05}=\cdots=N_{26,05}$.

Eq.(5.3$e$), particularly when $\epsilon_x\cdot\epsilon_{\ell x}=-1$, has
significance both for simplifying computer searches as in [19], and for
theoretical considerations (we will give one at the end of this section).

We will complete this general discussion with two theorems.
In the following, we will repeatedly make use of the fact that, if $N_1,N_2$
are the coefficient matrices of invariants of type $\T$, then so will be
$$aN_1+bN_2,\sp N_1^T,\sp N_1^{\dag},\sp N_1\,N_2$$
for any complex numbers $a,b$.

Call a (not necessarily physical)
invariant {\it 0-decoupled} if $N_{\la 0}=N_{0\la}=0$ for all $\la\ne 0$ in
$P_{+}$. A 0-decoupled invariant $Z$ can be written in the form
$$Z=a|\c_0|^2+Z',$$
where $Z'$ is independent of $\c_0$ and $\C_0$, and $a$ is any constant.
For example, the only 0-decoupled physical invariants for
$g=A_1$ are ${\cal A}_k$ $\forall k$, and those lying in the ${\cal D}_k$
series
with level $k\equiv 2$ (mod 4) (the $A_1$ physical invariants are given in
[10]). For $g=A_2$, ${\cal D}_k$ is 0-decoupled
$\forall
k \not\equiv 0$ (mod 3) (the known $A_2$ physical invariants are given \eg in
[14]).

By a {\it permutation invariant} we mean an invariant
$$Z=\sum_{\la\in P_{+}}\c_\la \,\c_{\sigma\la}^*,\sp i.e.\sp N_{\la\la'}=
\delta_{\la',\sigma\la}$$
for some permutation $\sigma$ of $P_{+}$. Any permutation invariant is
necessarily physical (see the proof of Thm.3 below).
Of course, the permutation $\sigma$ must be a symmetry of both the $S$ and $T$
matrices, and by Verlinde's formula [21], it also is a symmetry of the fusion
coefficients:
$$S_{\la\la'}=S_{\sigma\la,\sigma\la'},\sp T_{\la\la'}=
T_{\sigma\la,\sigma\la'},\sp N^{\la''}_{\la\la'}=
N_{\sigma\la,\sigma\la'}^{\sigma\la''}.$$

\bigskip \noindent{\bf Theorem 3}:\quad $N$ is a 0-decoupled physical invariant
iff it is a permutation invariant.

\bigskip \noindent{\it Proof}\quad Since a permutation invariant $N$
is necessarily
positive, it must satisfy $N_{00}>0$ by an observation made earlier in this
section. Hence $N_{00}=1$, $N$ is physical, and $N$ must be 0-decoupled.

Now consider $\tilde{N}=N\,N^T$, for some 0-decoupled physical invariant $N$.
$\tilde{N}$ will
also correspond to a 0-decoupled physical invariant, as will any of its powers
$(\tilde{N})^\ell$. Also, $$\tilde{N}_{\la\la}=\sum_{\la'}N_{\la \la'}^2 \ge
\sum_{\la'}N_{\la\la'} \equi r_\la \ge 0.$$

Suppose for contradiction that the row sum $r_\la>1$ for some $\la$, and
look at the powers $(\tilde{N})^\ell$. An easy calculation shows that
$(\tilde{N}^\ell)_{\la\la}\rightarrow \infty$ as $\ell\rightarrow \infty$,
contradicting (5.2). Therefore every entry in the $\la$-row is 0, except
for at most one 1.

A similar argument applies to columns, by considering $N^TN$. Hence
$\tilde{N}$ is diagonal, with 1's and 0's on the diagonal. But
$$1=\tilde{N}_{00}=\sum_{\la,\la'} S_{\la 0}S_{\la' 0}\tilde{N}_{\la\la'}=
\sum_{\la}S_{\la 0}^2 \tilde{N}_{\la\la}\le \sum_{\la}S_{\la}^2=1$$
and each $S_{\la 0}>0$, so equality can hold only if each $\tilde{N}_{\la\la}
=1$. \qquad QED
\bigskip

By a {\it block} we mean something of the form $m|\c_{\la_1}+\cdots+
\c_{\la_\ell}|^2$. $m$ is called the {\it scale}, and $\ell$ the {\it length}.
We will demand $m,\ell\neq 0$. Call an invariant $Z$ a {\it block-diagonal}
if it is the sum of {\it pairwise disjoint} blocks. For example, the diagonal
invariant is always a block-diagonal, with each block having length $\ell_i
=1$ and scale $m_i=1$. In fact, it is the only physical invariant which is both
0-decoupled and block-diagonal. The block-diagonal physical invariants of
$g=A_1$ are the diagonal invariants, along with the invariants in the
${\cal D}_k$
series with level $k\equiv 0$ (mod 4), and the exceptional invariants
${\cal E}_6$ and ${\cal E}_8$. For $g=A_2$, the block-diagonal invariants
 include ${\cal D}_k$ for $k\equiv 0$ (mod 3), as well as exceptional
invariants
of level $k=5,9$ and 21.

Most of the known physical invariants seem to be either permutation invariants,
block-diagonal, or products of two  such invariants.

\bigskip \noindent{\bf Theorem 4}:\quad Let $Z$ be any invariant (not
necessarily physical) which is a block-diagonal, and let the $i$th
block have scale $m_i$ and length $\ell_i$. Then:

\item{(i)}\quad $\c_0$ must belong to some block;

\item{(ii)}\quad all products $|m_i|\ell_i$ are equal;

\item{(iii)}\quad all $m_i$ must be of equal sign;

\item{(iv)}\quad if $Z$ is also {\it physical}, then each $\ell_i$ must divide
the length of the block containing $\c_0$.
\bigskip

\noindent{\it Proof}\quad The proof here is similar to that used in proving
the previous theorem; the key idea is that the coefficient matrix $N$ of
$Z$, after a re-ordering of the indices, is a direct sum of $\ell_i\times
\ell_i$ matrices looking like
$$m_i \left(\matrix{1&\cdots&1\cr\vdots&&\vdots\cr 1&\cdots&1\cr}\right)$$
with $\ell_i$ 1's in each row and column. Raising $N$ to the $\ell$-th
power then gives a direct sum of
$$m_i^\ell \ell_i^{\ell-1} \left(\matrix{1&\cdots&1\cr\vdots&&\vdots\cr 1&
\cdots&1\cr}\right). $$

To prove (i), simply square $N$ and use the fact that $N_{00}$ must be
positive.

To prove (ii), let $L$= max${}_i \,\,|m_i|\ell_i$ and look at the limit
$N_\infty$ of $\left( N^2/L^2 \right)^\ell$ as $\ell\rightarrow \infty$.
The only blocks that will have survived would be those with
$L^2=m_i^2\ell_i^2$. If $N_\infty\neq N^2/L^2$, then either $N_\infty$
or $N^2/L^2-N_\infty$ will be a block-diagonal positive invariant with
no block containing $\c_0$. Therefore $N_\infty$ must equal $N^2/L^2$.

(iii) follows from a similar argument, by looking at $N^2/L\pm N$.

(iv) follows immediately from (ii). \qquad QED

\bigskip Obviously Thms.3,4 are two among many that can be proved
using similar techniques. Thm.3 will play an important role in the complete
classification of the level 1 physical invariants given in Thm.5.

The remainder of this section will be devoted to the level
1 physical invariants, and in particular, the proof of the following theorem.

We will use the conventions for numbering Dynkin nodes given in Table Fin
in [3]. In eqs.(5.5$b$-$p$) we will write $\c_i$ for $\c_{\beta_i}(\T)$.

\bigskip \noindent{\bf Theorem 5}:\quad (i) For $g=B_n$, $E_7$, $E_8$, $F_4$
and $G_2$, the only level 1 physical invariant is the {\it diagonal invariant}:
$$Z_A=\sum_{\la\in P_{+}(g,1)}\c_\la^{g,1} \,\c_\la^{g,1*}. \eqno(5.5a)$$

\item{(ii)} For $g=E_6$, $k=1$, the physical invariants are the diagonal one
(5.5$a$), along with
$$Z_\sigma=\c_0\,\C_0+\c_1\,\C_5+\c_5\,\C_1.\eqno(5.5b)$$

\item{(iii)} For $g=C_n$, $k=1$, the physical invariants are: the diagonal one
(5.5$a$) for each $n$; for $n\equiv 0$ (mod 4) the invariant
$$Z_n'=2|\c_{n/2}|^2+\sum_{i=0}^{n/4-1}|\c_{2i}+\c_{n-2i}|^2;\eqno(5.5c)$$
for $n\equiv 2$ (mod 4), $n\ge 6$, the invariant
$$Z_n'=|\c_{n/2}|^2+\sum_{i=0}^{n/2}|\c_{2i}|^2+\sum_{i=0}^{n-6\over 4}
(\c_{2i+1}\,\c_{n-2i-1}^*+\c_{n-2i-1}\,\c_{2i+1}^*);\eqno(5.5d)$$
for $n=10,16,28$, respectively, the additional invariants
$$\eqalignno{Z_{10}''=&|\c_0+\c_6|^2+|\c_3+\c_7|^2+|\c_4+\c_{10}|^2;&(5.5e)\cr
Z_{16}''=&|\c_0+\c_{16}|^2+|\c_4+\c_{12}|^2+|\c_6+\c_{10}|^2+|\c_8|^2+&\cr
&\c_8\,(\c_2^*+\c_{14}^*)+(\c_2+\c_{14})\,\c_8^*;&(5.5f)\cr
Z_{28}''=&|\c_0+\c_{10}+\c_{18}+\c_{28}|^2+|\c_6+\c_{12}+\c_{16}+\c_{22}|^2.
&(5.5g)\cr}$$

\item{(iv)} For $g=D_n$, $k=1$, the physical invariants are: for $n\not\equiv
0$ (mod 4), the diagonal invariant (5.5$a$) along with
$$Z_\sigma=\c_0\C_0+\c_1\C_1+\c_{n-1}\C_n+\c_n\C_{n-1};\eqno(5.5h)$$
for $n\equiv 4$ (mod 8) the physical invariants are given by eqs.(5.5$a,h$)
and
$$\eqalignno{Z_3=&\c_0\C_0+\c_1\C_{n-1}+\c_{n-1}\C_1+\c_n\C_n,&(5.5i)\cr
Z_4=&\c_0\C_0+\c_1\C_{n}+\c_{n-1}\C_{n-1}+\c_n\C_1,&(5.5j)\cr
Z_5=&\c_0\C_0+\c_1\C_{n-1}+\c_{n-1}\C_n+\c_n\C_1,&(5.5k)\cr
Z_6=&\c_0\C_0+\c_1\C_n+\c_{n-1}\C_1+\c_n\C_{n-1};&(5.5l)\cr}$$
and for $n\equiv 0$ (mod 8) the physical invariants are given by eqs.(5.5$a,h$)
and
$$\eqalignno{Z_3'=&\c_0\C_0+\c_0\C_{n-1}+\c_{n-1}\C_0+\c_{n-1}\C_{n-1},
&(5.5m)\cr
Z_4'=&\c_0\C_0+\c_0\C_{n}+\c_{n}\C_{0}+\c_n\C_n,&(5.5n)\cr
Z_5'=&\c_0\C_0+\c_0\C_{n-1}+\c_{n}\C_0+\c_n\C_{n-1},&(5.5o)\cr
Z_6'=&\c_0\C_0+\c_0\C_n+\c_{n-1}\C_0+\c_{n-1}\C_{n}.&(5.5p)\cr}$$
\bigskip

Thus, together with the $g=A_n$, $k=1$ case dealt with in [11], this
theorem completes the classification of the level 1 physical invariants of
simple type. All the invariants in eqs.(5.5) are either 0-decoupled,
or diagonal-blocks, or products of the two, except for (5.5$f$).
Note further that not all of the physical invariants given in (5.5)
are {\it real} --- \ie correspond to symmetric coefficient matrices $N$.
In particular, each $g=D_{4\ell}$ has exactly two non-real level 1
physical invariants ($Z_5$ and $Z_6$, or $Z_5'$ and $Z_6'$). However, in terms
 of {\it restricted}
characters $\c(0,\u)$, all invariants become symmetric.

\bigskip \noindent{\it Proof}\quad
(i) We will do $B_n$ here; the other algebras in (i) are at least as easy.
First, we must find $P_{+}(B_n,1)$. This is easy:
$$P_{+}(B_n,1)=\{0,\beta_1,\beta_n\},$$
corresponding to the three colabels $a^{\vee}_i$ equal to 1. Therefore
any invariant will look like $Z=\sum N_{ij}\c_i\C_j$, where the sum is
over $i,j\in \{0,1,n\}$.

Invariance under $\u\rightarrow \u+1$ requires that we compute the norms
of $p_0\equi \rho/\sqrt{\h+1}$, $p_1\equi (\beta_1+\rho)/\sqrt{\h+1}$ and
$p_n\equi (\beta_n+\rho)/\sqrt{\h+1}$:
$$\eqalignno{p_0^2=&{4n^2-1\over 24},&(5.6a)\cr
p_1^2=&{4n^2+23\over 24}=p_0^2+1,&(5.6b)\cr
p_n^2=&{(n+1)(2n+1)\over 12}=p_0^2+{2n+1\over 4}. &(5.6c)\cr}$$
Eqs.(5.6) tell us that any invariant of type $(\{B_n,1\})$ looks like
$$Z=N_{00}\c_0\C_0+N_{11}\c_1\C_1+N_{nn}\c_n\C_n.\eqno(5.7)$$
But then any physical invariant of that type must be 0-decoupled. Thm.3
then tells us there must be exactly one 1 in each row and column, so
$N_{00}=N_{11}=N_{nn}=1$, and $Z=Z_A$, the diagonal invariant.

(ii) This case can be dealt with similarly. $P_{+}(E_6,1)=\{0,\beta_1,\beta_5
\}$; a norm check gives us
$$Z=N_{00}\c_0\C_0+\sum_{i,j\in\{1,5\}} N_{i,j}\c_i\C_j.\eqno(5.8)$$
Therefore any physical invariant of this type will necessarily be 0-decoupled,
so there are only two possibilities for these physical invariants:
eqs.(5.5$a,b$). Eq.(5.5$b$) is indeed an invariant;
it corresponds to the outer automorphism $\sigma$ of $E_6$.

(iii) The easiest completeness proof for $C_n$ is to exploit the calculation
in [22] of its level 1 $S^W$-matrix. An alternate argument, with the
advantage of greater generality, will be sketched at the end of this section.

$P_{+}(C_n,1)=\{0,\beta_1,\ldots,\beta_n\}$ and $P_{+}(A_1,k)=\{0,1\cdot
\beta_1',\ldots,k\cdot \beta_1'\}$, using obvious notation.
In [22] it is shown that
$$\bigl(S^W(C_n,1)\bigr)_{\beta_i\beta_j}=\bigl(S^W(A_1,n)
\bigr)_{i\beta_1',j\beta_1'},\eqno(5.9a)$$
for all $0\le i,j\le n$ ([23] generalized this duality to between $S^W(C_n
,k)$ and $S^W(C_k,n)$). To compare their $T^W$-matrices, we need to compute
the norms of $p_i\equi (\beta_i+\rho)/\sqrt{\h+1}$. We find
$$p_i^2={(n+1)(2n+3)\over 12}-{(n+1-i)^2\over 2(n+2)}.\eqno(5.9b)$$
Therefore from (2.6$b$) we get
$$\bigl(T^W(C_n,1)\bigr)_{\beta_i\beta_j}=\alpha\bigl(T^W(A_1,n)^{-1}
\bigr)_{i\beta_1',j\beta_1'},\eqno(5.9c)$$
for some $\alpha\in {\bf C}$ of modulus $|\alpha|=1$. From (5.9$a,c)$ we
see that the commutants of types $(C_n,1)$ and $(A_1,n)$ are the same:
more precisely, the map defined by $N_{\beta_i\beta_j}\rightarrow
N'_{i\beta_1',j\beta_1'}$ is a (vector space) isomorphism between
$\Omega_W(C_n,1)$ and $\Omega_W(A_1,n)$. Since it also preserves (P2) and
(P3), it constitutes a bijection between the respective physical invariants.

What this means is that the completeness proof for $A_1$ level $n$,
carries over to $C_n$ level 1, and the list of $C_n$ physical invariants
can be read off from that of $A_1$. The result is eqs.(5.5$c$-$g$).

(iv) Note that $P_{+}(D_n,1)=\{0,\beta_1,\beta_{n-1},\beta_n\}$, and the
norms of $p_0\equi \rho/\sqrt{\h+1}$ and $p_i\equi (\beta_i+\rho)/\sqrt{\h+1}$
for $i=1,n-1,n$ are:
$$p_0^2={n(n-1)\over 6},\sp p_1^2=p_0^2+1,\sp p_{n-1}^2=p_n^2=p_0^2+{n\over 4}
.\eqno(5.10)$$
Therefore, for $n\not\equiv 0$ (mod 4), $n\equiv 4$ (mod 8), and $n\equiv 0$
(mod 8) respectively, an invariant will look like
$$\eqalignno{Z&=N_{00}\c_0\,\C_0+N_{11}\c_1\,\C_1+N_{n-1,n-1}\c_{n-1}\,\C_{n-1}
+N_{nn}\c_n\,\C_n+N_{n-1,n}\c_{n-1}\,\C_n&\cr
&+N_{n,n-1}\c_n\,\C_{n-1},&(5.11a)\cr
Z&=N_{00}\c_0\,\C_0+N_{11}\c_1\,\C_1+N_{n-1,n-1}\c_{n-1}\,\C_{n-1}
+N_{nn}\c_n\,\C_n+N_{n-1,n}\c_{n-1}\,\C_n+&\cr &N_{n,n-1}\c_n\,\C_{n-1}
+N_{1,n-1}\c_1\,\C_{n-1}+N_{n-1,1}\c_{n-1}\,\C_1+N_{1n}\c_{1}\,\C_{n}
+N_{n1}\c_n\,\C_{1}&(5.11b)\cr
Z&=N_{00}\c_0\,\C_0+N_{11}\c_1\,\C_1+N_{n-1,n-1}\c_{n-1}\,\C_{n-1}
+N_{nn}\c_n\,\C_n+N_{n-1,n}\c_{n-1}\,\C_n+&\cr
&N_{n,n-1}\c_n\,\C_{n-1}
+N_{0,n-1}\c_0\,\C_{n-1}+N_{n-1,0}\c_{n-1}\,\C_0+N_{0n}\c_{0}\,\C_{n}
+N_{n0}\c_n\,\C_{0}&(5.11c)\cr}$$

Then any physical invariant for $n\not\equiv 0$ (mod 8) will be 0-decoupled, so
by Thm.3 the only possibilities for those invariants are
given in eqs.(5.5$a,h$-$l$). That all these invariants are realized can be
most easily seen using the list of physical invariants compiled in \eg [9].
(5.5$h$) is his invariant ${\hat M}^{[\mu_v]}$, while (5.5$i$) is his
${\hat M}^{[\mu_s]}$. Eqs.(5.5$j,k,l$) can now be obtained by multiplication:
\eg (5.5$i$) multiplied on the right and left by (5.5$h$) gives (5.5$j$).

That completes the proof for $n\not\equiv 0$ (mod 8). The case $n\equiv 0$
(mod 8) is more difficult. The following argument is only one of
many that can be used.

As before, [9] can be used to show (5.5$m$-$p$) are all invariants.
(Incidently,
[9] missed 2 of the physical invariants for each $n\equiv 0$ (mod 4).) That
(5.5$a,h$) are the only permutation invariants is easy to see from (5.11$c$).
So we may assume $N$ is non-0-decoupled.

Using (5.1) we can easily show $S^W_{00}=S^W_{01}=S^W_{0,n-1}=S^W_{0n}$;
since their squares must sum to 1 we know they must all equal ${1\over 2}$.
Then we can read off from (5.2) that
$$4N_{00}=\sum_{i,j=0,1,n-1,n}N_{ij}\eqno(5.12a)$$
for any invariant $N$.

Let $N$ be any non-0-decoupled physical invariant. By (5.12$a$),
$\sum N_{ij}=4$. Now, multiplying $N$ on the right
by the coefficient matrix of (5.5$m$) and using (5.12$a$) tells us
$$N_{00}+N_{0,n-1}=N_{n-1,0}+N_{n-1,n-1}+N_{n,0}+N_{n,n-1}.\eqno(5.12b)$$
Similar calculations give formulas for $N_{00}+N_{0,n}$, etc. From these
it follows that one of $N_{0,n-1}$, $N_{0n}$ must be 0 and the other 1, and
similarly for $N_{n-1,0}$, $N_{n0}$. Each one of the resulting 4 possible
assignments of 0,1 to $N_{0,n-1}$, $N_{0n}$, $N_{n-1,0}$ and
$N_{n0}$ turns out to fix all other $N_{ij}$, and gives us one of $(5.5m$-$p$).
\qquad QED\bigskip

The level 1 cases are sufficiently simple that several different proofs are
possible. An example presumably is an explicit computation of the commutant
based on the calculations of level 1 $S^W$-matrices made in [22]. The above
arguments have the big advantage of being applicable to the more complicated
types --- \eg $(A_2,k)$ (see [24]). An important tool
in these cases is (5.3$e$) and
the consequence for positive invariants when $\epsilon_x \cdot \epsilon_{\ell
x}
=-1$. To illustrate this, an alternate proof of $(C_n,1)$ will now be
provided for $n$ odd. Of course, it can also be used to prove
completeness for $A_1$, odd levels.

Our goal is to prove that the diagonal invariant is the only physical invariant
of these levels. We will accomplish this by first showing that the only
permutation invariant is the diagonal invariant. Then we will use (5.3$e$)
to show that any physical invariant must necessarily be 0-decoupled. Thm.3
then completes the argument.

Let $S_{ij}=S^W(C_n,1)_{\beta_i\beta_j}$. We get from (5.1), by computing
$S_{0,\ell+1}/S_{0\ell}$, that
$$S_{0\ell}=\beta_n\,{\rm sin}\,\pi{\ell+1\over n+2} \eqno(5.13a)$$
for some constant $\beta_n$. Let $N_{ij}=\delta_{j,\sigma i}$ be a
{\it permutation
invariant}. Then (2.7$b$) tells us $S_{ij}=S_{\sigma i,\sigma j}$. Since
$\sigma(0)=0$, (5.13$a$) now implies
$$\sigma(i)\in \{i,n-i\}.\eqno(5.13b)$$
But $(2.7a)$ tells us $T_{ij}=T_{\sigma i, \sigma j}$, \ie $p_i^2\equiv
p_{\sigma i}^2$ (mod 2). These norms are computed in (5.9$b$).
For odd $n$, $(n+1-i)^2\equiv (i+1)^2$ (mod $4(n+2)$) has {\it no}
solutions for $i$, so for $n$ odd, $\sigma(i)=i$ and the only permutation
invariant is the diagonal invariant (5.5$a$).

Now let $N_{ij}$ be any {\it positive invariant}. Suppose $N_{0k}>0$. We want
to show this can only happen when $k=0$. By $[m]$ we will mean the unique
number congruent to $m$ (mod $2(n+2)$) lying in $0\le [m]<2(n+2)$. Let
$x=(p_0;p_k)$. Then $\epsilon_x=+1$ (see (5.3$c$)). Let $\ell$ be relatively
prime to $2(n+2)=$ the order of $x$. Then a simple calculation we will not
include here shows
$$\eqalignno{\epsilon_{\ell x}=&\bigl\{(-1)^{(n+1-0)(\ell+1)}\epsilon'_{0,\ell}
\epsilon_\ell''\bigr\}\cdot \bigl\{(-1)^{(n+1-k)(\ell+1)}\epsilon'_{k,\ell}
\epsilon_\ell''\bigr\}=\epsilon'_{0,\ell}\cdot\epsilon'_{k,\ell},&(5.14a)\cr
{\rm where}&\sp \epsilon'_{i,j}=\left\{\matrix{+1&{\rm if}\sp 0<
[(n+1-i)j]<n+2\cr -1&{\rm if}\sp [(n+1-i)j]>n+2\cr}\right. ,&(5.14b)\cr}$$
and $\epsilon''_\ell$ is a sign depending only on $\ell$.
Then $(5.3e)$, together with $N_{ij}$ being a positive invariant, and $N_{0k}
>0$, forces $\epsilon'_{0,\ell}=\epsilon'_{k,\ell}$.

Consider $\ell_m=2^m+n+2$, for $m>0$. Each $\ell_m$ is relatively prime to
$2(n+2)$, since $n+2$ is odd. $(2.7a)$ implies $x^2\equiv 0$ (mod 2), which
by $(5.9b)$ implies $k$ must be even. Therefore $\epsilon'_{0,\ell_m}=
\epsilon'_{0,2^m}$ and $\epsilon'_{k,\ell_m}=\epsilon'_{k,2^m}$. Now let
$M$ be defined by $2^M<n+2<2^{M+1}$. Then $\epsilon'_{0,2^m}=-1$ $\forall
m=1,\ldots,M$. Hence $\epsilon'_{k,2^m}=-1$ for those $m$. Since in addition
$0<n+1-k<n+2$, these force $n+1-k=n+1$, \ie $k=0$.

Similarly, $N_{k0}>0$ would also imply $k=0$. Therefore the only positive
invariants for odd $n$ are 0-decoupled. This completes the proof that, for
odd $n$, the only {\it physical invariant} of $C_n$ level 1, or $A_1$
level $n$, is the diagonal invariant.

\bigskip\bigskip \centerline{{\bf 6. Conclusion}} \bigskip

We begin the paper by generalizing the analysis of [14] to any semi-simple
algebra. We then prove that the Roberts-Terao-Warner lattice method is
complete, and make the method more efficient by limiting the class of lattices
that need be considered. By making use of the additional multiplicative
structure
the commutant has, which was not exploited by [14], we obtain some
results in Sec.5 which help us to completely  classify all level 1
physical invariants --- see eqs.(5.5). The most attractive feature of the
arguments in Sec.5 is their generality.

 For small ranks and levels, the lattice method for generating invariants
is computationally speaking quite
practical, as is shown by the work in [19]. But its greatest value may be
that it provides a convenient theoretical description for the commutant.
A natural first step
for classifying all physical invariants in a given class (see \eg [10,14])
involves understanding the commutant, and lattices could provide a valuable
geometrical tool for that.
An exciting recent developement was the translation given in [25] of the
$A_1$ completeness proof into the lattice language (another proof, for odd
levels, is included at the end of Sec.5). The resulting argument
was surprisingly simple, suggesting that the Roberts-Terao-Warner approach
may be a particularly fruitful one for the search for the other, more elusive
completeness proofs (\eg $A_2$). This direction is being actively pursued in
[24]. Of course the proof in this paper that
lattice partition functions span the commutant, is a necessary first step
for this program.

A generalization of this lattice approach, for use in finding
 {\it heterotic} invariants, was given in [6]. Among other things, it
does not require the self-dual lattices to be even.
Included there was a proof
that it generates all heterotic invariants. Using this generalization, it
may also be possible to apply lattices to coset models (see [25]), but work
 in that direction has not yet been completed.

Another avenue suggested by this paper lies in investigating the structure
of the commutant as an algebra. More work in this direction is currently
underway.

\bigskip
This work is supported in part by the Natural Sciences and Engineering
Research Council of Canada. It grew out of an ongoing collaboration
with Quang Ho-Kim, C.S.{} Lam and Patrick Roberts; Dr. Roberts has been
particularly valuable as a proof-reader. I also appreciate the
hospitality shown by the Carleton mathematics department, where this paper
was written.

\bigskip \bigskip\centerline{{\bf References}} \bigskip

\item{[1]} A.A. Belavin, A.M. Polyakov and A.B. Zamolodchikov, {\it Nucl.
Phys.} {\bf B241} (1984) 333

\item{[2]} D. Gepner and E. Witten, {\it Nucl. Phys.} {\bf B278} (1986) 493

\item{[3]} V.G. Kac, {\it Infinite Dimensional Lie Algebras}, 3rd ed.
(Cambridge University Press, Cambridge, 1990)

\item{[4]} S. Kass, R.V. Moody, J. Patera and R.Slansky, {\it Affine Lie
Algebras, Weight Multiplicities, and Branching rules} Vol.1 (University of
California Press, Berkeley, 1990)

\item{[5]} A.N. Schellekens, {\it Classification of ten-dimensional
heterotic strings} (CERN preprint TH.6325, 1991)

\item{[6]} T. Gannon, {\it Partition functions for heterotic WZW
 conformal field theories} (Carleton preprint, 1992)

\item{[7]} S. Bais and P. Bouwknegt, {\it Nucl. Phys.} {\bf B279} (1987) 561;

\item{} A.N. Schellekens and N.P. Warner, {\it Phys. Rev.} {\bf D34} (1986)
3092

\item{[8]} A.N. Schellekens and S. Yankielowicz, S. {\it Nucl. Phys.}
 {\bf B327} (1989) 673

\item{[9]} D. Bernard, {\it Nucl. Phys.} {\bf B288} (1987) 628

\item{[10]} A. Cappelli, C. Itzykson and J.-B. Zuber, {\it Nucl. Phys.}
{\bf B280 [FS18]} (1987) 445;

\item{} A. Cappelli, C. Itzykson and J.-B. Zuber,
{\it Commun. Math. Phys.} {\bf 113} (1987) 1;

\item{} D. Gepner and Z. Qui, {\it Nucl. Phys.} {\bf B285} (1987) 423

\item{[11]} C. Itzykson, {\it Nucl. Phys. (Proc. Suppl.)} {\bf 5B} (1988) 150

\item{} P. Degiovanni {\it Commun. Math. Phys.} {\bf 127} (1990) 71

\item{[12]} N.P. Warner, {\it Commun. Math. Phys.} {\bf 130} (1990) 205

\item{[13]} P. Roberts and H. Terao, {\it Int. J. Mod. Phys.} {\bf A7} (1992)
2207

\item{[14]} M. Bauer and C. Itzykson,
  {\it Commun. Math. Phys.} {\bf 127} (1990) 617

\item{[15]} J.H. Conway and N.J.A. Sloane, {\it Sphere packings,
Lattices and Groups} (Springer-Verlag, New York, 1988)

\item{[16]} B. Schoenberg, {\it Elliptic Modular Functions} (Springer-Verlag,
Berlin, 1974)

\item{[17]} Ph. Ruelle, {\it Commun. Math. Phys.} {\bf 133} (1990) 181

\item{[18]} T. Gannon and C.S. Lam, {\it Rev. Math. Phys}. {\bf 3} (1991) 331

\item{[19]} Q. Ho-Kim and T. Gannon, {\it The low level modular invariants
of rank 2 algebras} (work in progress)

\item{[20]} Ph. Ruelle, E. Thiran and J. Weyers, {\it Comm. Math. Phys.}
{\bf 133} (1990) 305

\item{[21]} E. Verlinde, {\it Nucl. Phys.} {\bf B300 [FS22]} (1988) 360;

\item{} G. Moore and N. Seiberg, {\it Phys. Lett.} {\bf B212} (1988) 451

\item{[22]} V.G. Kac and M. Wakimoto, {\it Adv. Math.} {\bf 70} (1988) 156

\item{[23]} D. Verstegen, {\it Comm. Math. Phys.} {\bf 137} (1991) 567

\item{[24]} T. Gannon and P. Roberts, {\it The modular invariant partition
functions of SU(3)} (work in progress)

\item{[25]} P. Roberts, {\it Whatever Goes Around Comes Around:
Modular Invariance in String Theory and Conformal Field Theory}, Ph.D. Thesis
(Institute of Theoretical Physics, Goteborg, 1992)

\end